\documentclass[12pt,fleqn]{article} 

\usepackage{graphicx}
\usepackage{dcolumn}
\usepackage{bm}
\usepackage{amsmath}
\usepackage{amssymb}  

\def\OMIT#1{}
\begin{document}

\title{Transcription and noise in negative feedback loops} 
\author{J.C. Nacher$^1$ and T. Ochiai$^2$}

\maketitle
\begin{center}
{\it $^1$ Department of Complex Systems, Future University-Hakodate}
\end{center}
\begin{center}
{\it 116-2 Kamedankano Hakodate, Hokkaido, 041-8655, Japan}
\end{center}
\begin{center}
nacher@fun.ac.jp
\end{center}
\begin{center}
{\it $^2$ Faculty of Engineering, Toyama Prefectural University }
\end{center}
\begin{center}
{\it 5180 Kurokawa Imizu-shi Toyama, 939-0398, Japan}
\end{center}
\begin{center}
ochiai@pu-toyama.ac.jp
\end{center}

\maketitle
 
\begin{center}
\end{center} 

\begin{center}

\end{center} 

\begin{abstract}
{\small{
Recently, several studies have investigated the transcription process associated to specific genetic regulatory networks.
In this work, we present a stochastic approach for analyzing the dynamics and effect of negative feedback loops (FBL) 
on the transcriptional noise. First, our analysis allows us to identify a bimodal activity depending of the strength of 
self-repression coupling $D$. In the 
strong coupling region $D>>1$, the variance of the 
transcriptional noise is found to be reduced a 28 $\%$ more than described 
earlier. Secondly, the contribution of the noise effect to the
abundance of regulating protein becomes manifest when the coefficient of variation is computed. In the strong coupling region, 
this coefficient is found to be independent 
of all parameters and in fair agreement with the experimentally observed values. Finally, our analysis reveals that 
the regulating protein is significantly induced by the intrinsic and external noise in the strong coupling region. 
In short, it indicates that the existence of inherent noise in FBL makes it possible 
to produce a basal amount of proteins even though the repression level $D$ is very strong. 

}}

\end{abstract}

\section{Introduction}

The cell is a highly dynamic and regulated system composed of complex pathways and networks formed by tens thousands of inter-connected proteins, genes, and metabolites. Gene expression regulation is a complex cellular process that involves different genetic elements, through which cells control multiple functions such as the synthesis of mRNA molecules and the production of enzymatic proteins. Recently, it has been shown that stochastic fluctuations in populations of genetic 
and biochemical molecules can influence the gene regulatory processes \cite{kaern, swain2002, kepler2001}. Each cell represents 
a complex system that has evolved in the presence of considerable variations and random fluctuations 
of molecular components. As a consequence, cells have been adapted to exploit the noise to enhance 
cellular processes \cite{chen2005,gardner,paulsson2004,elowitz2002,austin2006,blake2003,raser2004}.

In the transcriptional process, a part of DNA sequence (gene) is copied by an RNA polymerase to 
synthesize mRNA molecule. In a second step, mRNA is decoded to produce specific gene products 
like transcriptional factors (TF) or proteins. This transcriptional process can be affected 
by two sources of noise. While the internal noise emerges from low copy number of molecules 
and the random encounters between reactants, the external noise is related to changes 
in the neighborhood and environmental conditions \cite{swain2002, paulsson2004}.

One mechanism that the cell uses to deal with noise is the negative feedback loop (FBL) \cite{savageau1974}. Both 
theoretical and experimental studies on a FBL in Escherichia coli showed that these auto-regulatory genetic 
modules decrease transcriptional noise and enhance the stability in genetic networks 
\cite{paulsson2004,austin2006,becskei,rosenfeld2002}. Furthermore, the propagation of noise 
in genetic networks is a further interesting question \cite{pedraza2005, hooshangi2005}. However, challenges 
still remain to obtain a more accurate description of the transcriptional process governed by negative feedback loops.

In this paper, we study the auto-regulatory genetic module using a stochastic model. The original work that 
experimentally and theoretically
analyzed this module was shown in \cite{becskei}. We first derive the potential and the 
gene product probability distributions corresponding to the
auto-regulatory module. We then evaluate the role of noise by computing the expectation and variance values. We show that 
our approach leads
to new insights into the FBL. We were able to characterize the system according to two different phases: weak and strong feedback regions.
 In particular, our model predicts that the FBL decreases noise in the strong coupling region a 28 $\%$ more than described earlier
\cite{becskei}. Furthermore, we obtained a coefficient of variation of 0.75 in the strong coupling region, which is independent of all
parameters. Remarkably, this predicted value is in agreement with the experimental value observed in \cite{serrano2}. 

The paper is organized as follows. First, we describe the formulation of our model. 
Next section shows the results classified in subsections corresponding to the potential, probability distribution, expectation value, 
variance and noise dependence. The last section discusses and summarizes our findings.

\section{Stochastic model formulation}

A deterministic model to study the auto-regulatory module was first introduced by \cite{Denise}. The model was based on 
thermodynamic theory and kinetics associated to the system (see \cite{Denise} for details). The RNA polymerase 
is possibly bound to the promoter and the protein is also bound to the operator site. Therefore, the single gene, single 
promoter and single operator site system has three different configurations. The first state (s=1) 
is that neither the RNA polymerase nor protein are bounded to the promoter and the operator site, respectively. The second 
state (s=2) is that the RNA polymerase is bound to the promoter, but the protein is not bound to the operator. The third 
state (s=3) is that the RNA polymerase is not bound to the promoter, but the protein is bound to the operator. The state 
that both the RNA polymerase and protein are bound to the promoter and the operator site respectively is prohibited, since 
repressor protein prevents the RNA polymerase from attaching the promoter site. According to the model, the 
concentration $x$ of unbounded regulating proteins obeys the following equation:

\begin{eqnarray}\label{eqn: regulation equation}
\frac{dx}{dt}=\frac{k \alpha e^{-\Delta G_2/RT}[RNAP]}{e^{-\Delta G_1/RT}+e^{-\Delta G_2/RT}[RNAP]+e^{-\Delta 
G_3/RT}x}-\lambda x,
\end{eqnarray}
where $\Delta G_s$ is the Gibbs free energy of state $s$, $R$ is the gas constant, $T$ is the absolute temperature, 
$[RNAP]$ is the concentration of unbound RNA polymerase molecules, $x$ is the concentration of unbounded 
regulating protein, $k$ is the rate of RNA polymerase isomerization from closed to open complex, $\alpha$ is 
the proportionality coefficient that represents the number of protein synthesized per complex formed, and $\lambda$ 
is the protein degradation rate. To clarify the argument, we simplify equation 
(\ref{eqn: regulation equation}) as follows:

\begin{eqnarray}
\frac{dx}{dt}=\frac{A}{C+Dx}-\lambda x,
\end{eqnarray}
where $A=k \alpha e^{-\Delta G_2/RT} [RNAP]$, $C=e^{-\Delta G_1/RT} + e^{-\Delta G_2/RT} [RNAP]$, $D=e^{-\Delta 
G_3/RT}$.

By using a stochastic partial differential equation (SPDE), we can derive a stochastic regulatory model 
that allows us to include the stochastic nature of the transcriptional process. This can be 
done by replacing the usual variable $x$ by the stochastic variable and adding the noise term in Eq. (1). As a result, 
the SPDE of one gene, one operator-site system is given by

\begin{eqnarray}\label{eqn: fundamental equation2}
dX_t&=&a(X_t)dt+\sigma dW_t,
\end{eqnarray}
where
\begin{eqnarray}
a(X_t)=\frac{A}{C+DX_t}-\lambda X_t.
\end{eqnarray}

Here, the stochastic variable $X_t$ denotes the fluctuating concentration of unbounded regulating protein, 
$W_t$ corresponds to the Wiener process and $\sigma$ represents the combined effect of internal and external noise.

\section{Results}

\subsection{Potential representation of the FBL}

The potential of the system given by Eq. (\ref{eqn: fundamental equation2}) can read as follows:
\begin{eqnarray}\label{eqn: potential}
U(x)&=&-\int^x a(s)ds \nonumber\\
&=&-\frac{A}{D}\log(C+Dx)+\frac{1}{2}\lambda x^2 ~~~~~(x>-\frac{C}{D}).
\end{eqnarray}

The strength of the coupling between the unbounded regulating proteins (also known as transcriptional factors) and the 
operator site is represented by the parameter $D$. Then, high $D$ values indicate strong coupling probability. Then, 
FBL will strongly repressed the production of new regulating proteins. In the following, we analyze 
how is the shape of the potential in both weak and strong coupling regions.  

\paragraph{The weak coupling limit}

In the weak feedback region $D \to 0$, the potential (\ref{eqn: potential}) takes the form:
\begin{eqnarray}\label{eqn: potential weak}
U(x)=\frac{1}{2}\lambda(x-\frac{A}{\lambda C})^2-\frac{A^2}{2\lambda C^2}-\frac{A}{D}\log C
\end{eqnarray}
This is the classical Gaussian potential.

\paragraph{The strong coupling limit}
In the strong feedback region $D \to \infty$, the shape of the potential (\ref{eqn: potential}) is transformed into the following expression:
\begin{eqnarray}\label{eqn: potential strong}
U(x)=\left \{ 
\begin{array}{l}
\frac{1}{2}\lambda x^2 ~~~~~(x>0) \\
\infty ~~~~~(x=0).
\end{array}
\right.
\end{eqnarray} 
This is a truncated like potential. Truncation naturally arises due to the log term in the potential (\ref{eqn: potential}). When $D$ 
increases, there is 
a transition from the Gaussian potential (\ref{eqn: potential weak}) to the truncated potential (\ref{eqn: potential strong}). The 
existence of two different shapes of potentials depending on the strength coupling $D$ has implications 
in the probability distributions as we will show in the next section.

\subsection{Probability Distribution}

It is known that SPDE's can be generally transformed into Fokker-Planck equations (FK equations) which are 
mathematically equivalent to the original SPDE (See \cite{Wong, Kampen, black} for details). Then, we 
can transform Eq. (\ref{eqn: fundamental equation2}) into
the following FK equation: 

\begin{eqnarray}\label{eqn: FK equation}
\frac{\partial p(x,t)}{\partial t}=\frac{\partial}{\partial 
x}\{U^\prime(x)p(x,t)\}
+\frac{\sigma^2 }{2}\frac{\partial^2}{\partial x^2}\{p(x,t)\}.
\end{eqnarray}
By solving this FK equation, the stationary distribution is given by
\begin{eqnarray}\label{eqn:stationary distribution}
p(x)&=&K \exp(-\frac{2}{\sigma^2}U(x)) \nonumber \\
&=&K \exp(\frac{2}{\sigma^2}\frac{A}{D}\log(C+Dx)-\frac{\lambda}{\sigma^2}x^2),
\end{eqnarray}
where $K$ is a normalization constant. Again, we can analyze the weak and strong coupling 
of unbounded transcriptional factors and gene operator site as follows.

\paragraph{The weak coupling limit}
In the weak feedback region, $D \to 0$, the probability distribution (\ref{eqn:stationary distribution}) reads as
\begin{eqnarray}
p(x)&=&\sqrt{\frac{\lambda}{\sigma^2\pi}} \exp(-\frac{\lambda}{\sigma^2}(x-\frac{A}{\lambda C})^2).
\end{eqnarray}
This is a Gaussian distribution.

\paragraph{The strong coupling limit}
In the strong feedback region $D \to \infty$, the distribution (\ref{eqn:stationary distribution}) changes and takes the form
\begin{eqnarray}
p(x)=\left \{ 
\begin{array}{l}
2\sqrt{\frac{\lambda}{\sigma^2\pi}} \exp(-\frac{\lambda}{\sigma^2}x^2) ~~~~~(x>0) \\
0 ~~~~~(x \le 0).
\end{array}
\right.
\end{eqnarray}
This distribution corresponds to the truncated potential (\ref{eqn: potential strong}). It is worth noticing that
this truncated distribution emerges as a consequence of the truncated potential (\ref{eqn: potential strong}).
Next, in order to investigate the role of noise in the FBL module, we compute the expectation and variance values in both coupling limits.

\subsection{Expectation value}
\begin{figure}[htb]
\setlength{\unitlength}{1cm}
\begin{picture}(15,12)(-1,-1)
\put(-1,0){\includegraphics[scale=0.5]{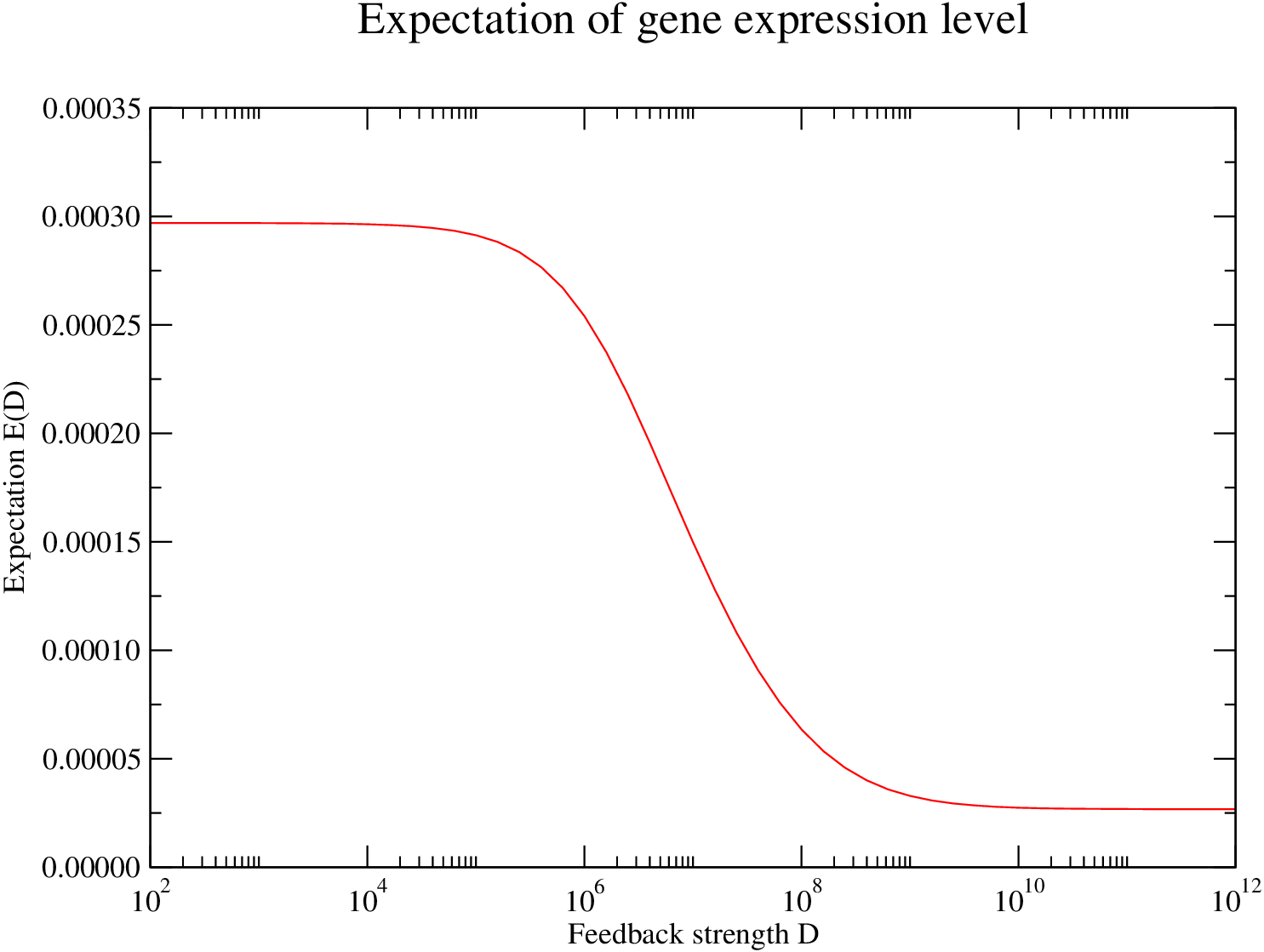}}
\end{picture} 
\caption{\small{
We show the expectation value of gene expression level in the auto-regulatory genetic module. Horizontal axis denotes feedback strength $D$ and vertical axis denotes the expectation value of gene expresssion level $E(D)$. The numerical values of the system parameters are taken from \cite{becskei}: $A=4.5\times 10^{-6} [\mbox{Ms}^{-1}], C=1.5\times 10^3, \lambda=10^{-5} [\mbox{s}^{-1}]$, and 
 $\sigma=1.5 \times 10^{-7}[\mbox{Ms}^{-1/2}]$ is an arbitrary parameter.}}
\label{fig: expectation}
\end{figure}

The expectation value of gene expression level in the auto regulatory module is given as
\begin{eqnarray}\label{eqn: expectation}
E(D)=<x>=\int_{-C/D}^\infty x p(x) dx.
\end{eqnarray}
The numerical solution of this expression is shown in Fig. \ref{fig: expectation}. We observe 
two different states depending on the feedback strength $D$. In particular, we see 
a transition from high to low expectation values with increasing the feedback 
strength $D$. While it is difficult to obtain the analytical expression for all $D$, we can 
derive these two states by computing Eq. (\ref{eqn: expectation}) in the coupling limits.

\paragraph{The weak coupling limit}
In the weak feedback region $D \to 0$, the expectation value (\ref{eqn: expectation})  is given by\footnote{Here we write $E(D=0)$ 
instead of E(0) for highlighting the variable D.} 
\begin{eqnarray}
E(D=0)=\frac{A}{\lambda C}.
\end{eqnarray}

\paragraph{The strong coupling limit}
In the strong feedback region $D\to \infty$, the expectation value (\ref{eqn: expectation})  reads
\begin{eqnarray}
E(D=\infty)=\frac{\sigma}{\sqrt{\lambda \pi}}.
\end{eqnarray}
This is the expectation value corresponding to the truncated potential. It is particularly 
clear on this result that, in the strong coupling limit, 
the abundance of unbounded regulating protein is caused
by the noise $\sigma$. In Fig. \ref{fig: expectation}, we see that even in 
the strong feedback coupling region $D$ (strongly self-repressed gene), there is a 
non-zero basal amount of proteins that emerges in our approach from 
the stochastic noise. In addition, it is worth noticing that deterministic analyses 
could not detect this protein concentration.

\subsection{Variance}
Next, we investigate the variance of gene expression level in FBL. Previous studies \cite{becskei} have shown 
that variance of gene expression is strongly reduced in FBL modules. However here, as a main result we find 
here that, in the strong coupling region, the feedback loop decreases the transcriptional noise in almost a 
30 $\%$ more than described in previous studies. First, the variance is defined as

\begin{eqnarray}\label{eqn: variance}
V(D)=<(x-E(D))^2>=\int_{-C/D}^\infty (x-E(D))^2 p(x) dx.
\end{eqnarray}
Eq. (\ref{eqn: variance}) was simulated and the result is shown in Fig. 2 in continuous line. In contrast, in dashed lines shown in Fig. 2, 
we can see the function corresponding to $1/S_r$ shown in \cite{becskei}. Furthermore, we analytically evaluate 
the variance for both limiting coupling regions.

\paragraph{The weak coupling limit}
In the weak feedback region $D \to 0$, the variance (\ref{eqn: variance}) reads as
\begin{eqnarray}
V(D=0)=\frac{\sigma^2}{2\lambda}.
\end{eqnarray}

\paragraph{The strong coupling limit}
In the strong feedback region $D\to \infty$, the variance (\ref{eqn: variance}) is given by
\begin{eqnarray}
V(D=\infty)=\frac{\sigma^2}{\lambda}(\frac{1}{2}-\frac{1}{\pi}).
\end{eqnarray}

\paragraph{Comparison with previous studies}
In \cite{becskei}, a ratio of the absolute values of stability of the unregulated to auto-regulated modules 
was used to compare both systems.
This is based on the fact that a system with higher stability exhibits a lower variance in gene expression. An 
equivalent computation can be
performed in our analysis by evaluating the ratio of variances at different coupling limits. In this case, a very weak coupling limit
($D\rightarrow 0$)
corresponds to the unregulated module considered in \cite{becskei}. The ratio is computed as follows:

\begin{eqnarray}
\frac{V(D=\infty)}{V(D=0)}=(1-\frac{2}{\pi})=0.363.
\end{eqnarray}
This expression is equivalent to the inverse of the relative stability $1/S_r$ (in the strong coupling region), 
shown in Eq. (3) of \cite{becskei}. In \cite{becskei}, it was found that this ratio takes the value 0.5. Therefore, 
we see that
our analysis suggests that the variance of the gene expression in the strong coupling region is reduced further than expected (28$\%$ more)
\cite{becskei}. This can also be seen by comparing the continuous and dashed line shown in Fig. 2.

\subsection{Coefficient of variation }

It is important to consider a magnitude that combines the expectation value and the variance. The coefficient 
of variation $C_V$ is a measure of dispersion of a probability distribution. It is useful for comparing the 
uncertainty between different measurements of varying absolute magnitude. It is defined 
as the ratio of the standard deviation to the mean. This magnitude is often called variability 
or relative standard deviation when the absolute value of the $C_V$ is expressed as a percentage. 
\begin{eqnarray}
C_V(D)=\frac{\sqrt{V(D)}}{E}
\end{eqnarray}
Fig.3 shows the result of the simulation of Eq. (19) for three sets of parameters.

\paragraph{The weak coupling limit}
In the weak feedback region $D \to 0$, the coefficient of variation reads as
\begin{eqnarray}
C_V(D=0)=\frac{\sqrt{\lambda}\sigma C}{\sqrt{2}A}.
\end{eqnarray}

\paragraph{The strong coupling limit}
In the strong feedback region $D\to \infty$, the coefficient of variation follows
\begin{eqnarray}
C_V(D=\infty)=\sqrt{\frac{\pi}{2}-1}=0.755.
\end{eqnarray}
We see that in the case of a very large feedback strength $D$, the coefficient of variation 
$C_V$ takes the value 0.75. Interestingly, this expression is independent of parameters. 
Fig. 3 shows the result of the simulation of Eq. (19) for three different sets of parameters. In 
all cases, we observed the same behaviour. In weak coupling limit ($D\to 0$), $C_V$ can be reduced by decreasing noise $\sigma$. However,
in the strong coupling limit, we cannot reduce $C_V$, even if we decrease the noise $\sigma$. This is because 
in the strong coupling limit, the
abundance of protein is linear in noise $\sigma$, which cancels the noise dependence coming from the 
standard deviation contribution. Therefore, 
in the strong coupling limit, the relative
value of variation $C_V$ can not be further reduced. In other words, no external parameters can disturb the system.Thus, a 
FBL system operating in the strong feedback region is robust. 

On the other hand, it is 
worth noticing that values in the vicinity of 0.75 were experimentally observed in \cite{serrano2}. Three different 
negative feedback loops
were designed and analyzed under different concentrations of anhydrotetracycline hydrochloride (aTc) ranging from 0 to 100 ng/ml. These
aTc molecules inhibit the negative feedback loop. These chemicals play the same role as 
the strength of regulation $D$ in our approach. At low aTc concentrations, the strength of the FBL is strong ($D$ large). In contrast,
high aTc concentrations correspond to low $D$ values. Experimental results described in \cite{serrano2} show that at very low aTc 
concentrations (i.e., $D\rightarrow \infty$), the $C_V$ is in the vicinity of 0.75, in agreement with our theoretical results.

\subsection{Noise dependence}
This system is characterized by two phases or regions depending on 
the strength of the self-repression coupling $D$. In strong coupling region, the role 
of noise is more relevant. We here address the issue of analytically 
assessing the role of noise in the FFL module. We define the 
following noise dependence $N$ \footnote{In (\ref{eqn: expectation}), we write $E(D)$ for expectation. However, in this section, we write $E(D,\sigma)$ for that, since the expectation also depends on $\sigma$.} 
\begin{eqnarray}
N(D)=\frac{E(D,\sigma)-E(D,0)}{E(D,\sigma)}.
\end{eqnarray}
This value $N(D)$ indicates the contribution of the noise effect $\sigma$ to the abundance (expectation value) of protein.
In the strong coupling region, $N(D)=1$, therefore it indicates 
that the system is completely dominated by the noise in this region. 
In contrast, in weak coupling region, $N(D)=0$. Then, the influence of the noise 
is very small in this limit. We showed the numerical solution in Fig. 4. 

\section{Conclusions}

There have been a series of studies and model developments in recent years towards an understanding of 
small functional units of cells. Among them, FBL is an interesting module with important 
implications for cell regulation and stability \cite{becskei}.

A major challenge addressed in this study consisted in embedding a stochastic approach 
into the structure of this FBL. This approach based on stochastic theory presents 
a number of advantages if compared with deterministic analyses. We summarize them as follows:
(1) We were able to identify a bimodal activity depending of the strength of self-repression coupling $D$. (2) In the 
strong coupling region, the variance of the 
transcriptional noise was found to be reduced a 28 $\%$ more than described 
earlier \cite{becskei}. (3) The contribution of the noise effect to the
abundance of regulating protein becomes manifest when the coefficient of variation was computed. This value was independent 
of all parameters and in fair agreement with the experimentally observed values \cite{serrano2}. This result could 
have not been found using deterministic models. (4) Our analysis revealed that 
the autoregulation process is significantly induced by the intrinsic and external noise in the strong coupling region. 
In short, it means that the existence of inherent noise in FBL makes it possible 
to produce a basal amount of proteins even although the repression level $D$ is very high. 

Finally, it remains to be explored to which extent this stochastic analysis can be extended to 
the analysis of noise propagation in networks composed of several modules \cite{pedraza2005, hooshangi2005}, and even 
in larger gene networks, and more interestingly in which way these large-scale networks can increase the robustness and stability.

\begin{figure}[htb]
\setlength{\unitlength}{1cm}
\begin{picture}(15,12)(-1,-1)
\put(-1,0){\includegraphics[scale=0.5]{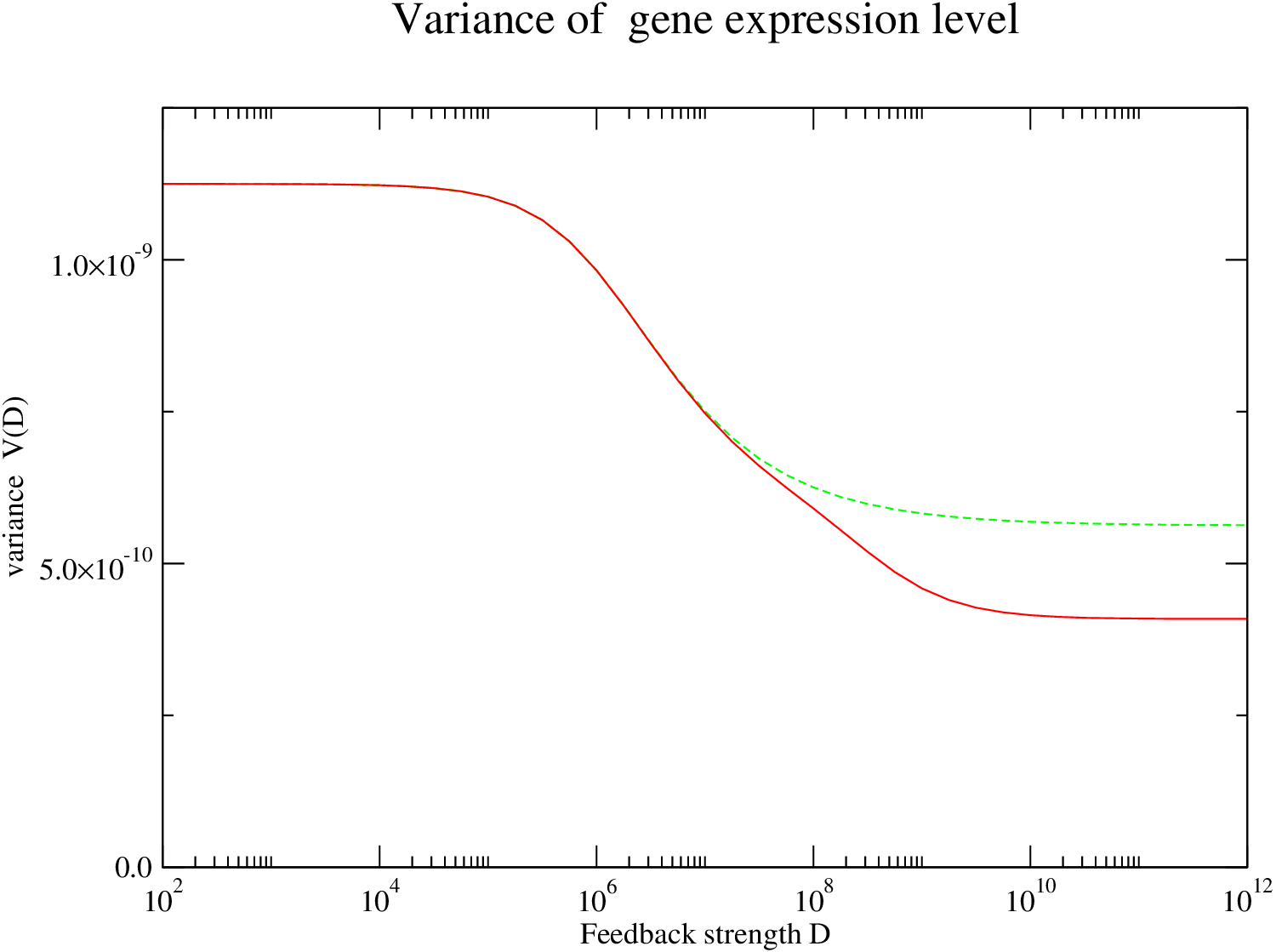}}
\end{picture} 
\caption{\small{
We show the variance of gene expression level in the auto-regulatory genetic module. Horizontal axis denotes feedback strength $D$ and vertical axis denotes the variance $V(D)$. The continuous line is our result and the dashed line is the result from \cite{becskei}. We see that the variance in the gene expression is reduced further than expected (28$\%$ more) in the strong coupling region. The numerical values of the system parameters are the same as Fig. \ref{fig: expectation}. }}
\label{fig: adjusting time}
\end{figure}

\begin{figure}[htb]
\setlength{\unitlength}{1cm}
\begin{picture}(15,12)(-1,-1)
\put(-1,0){\includegraphics[scale=0.5]{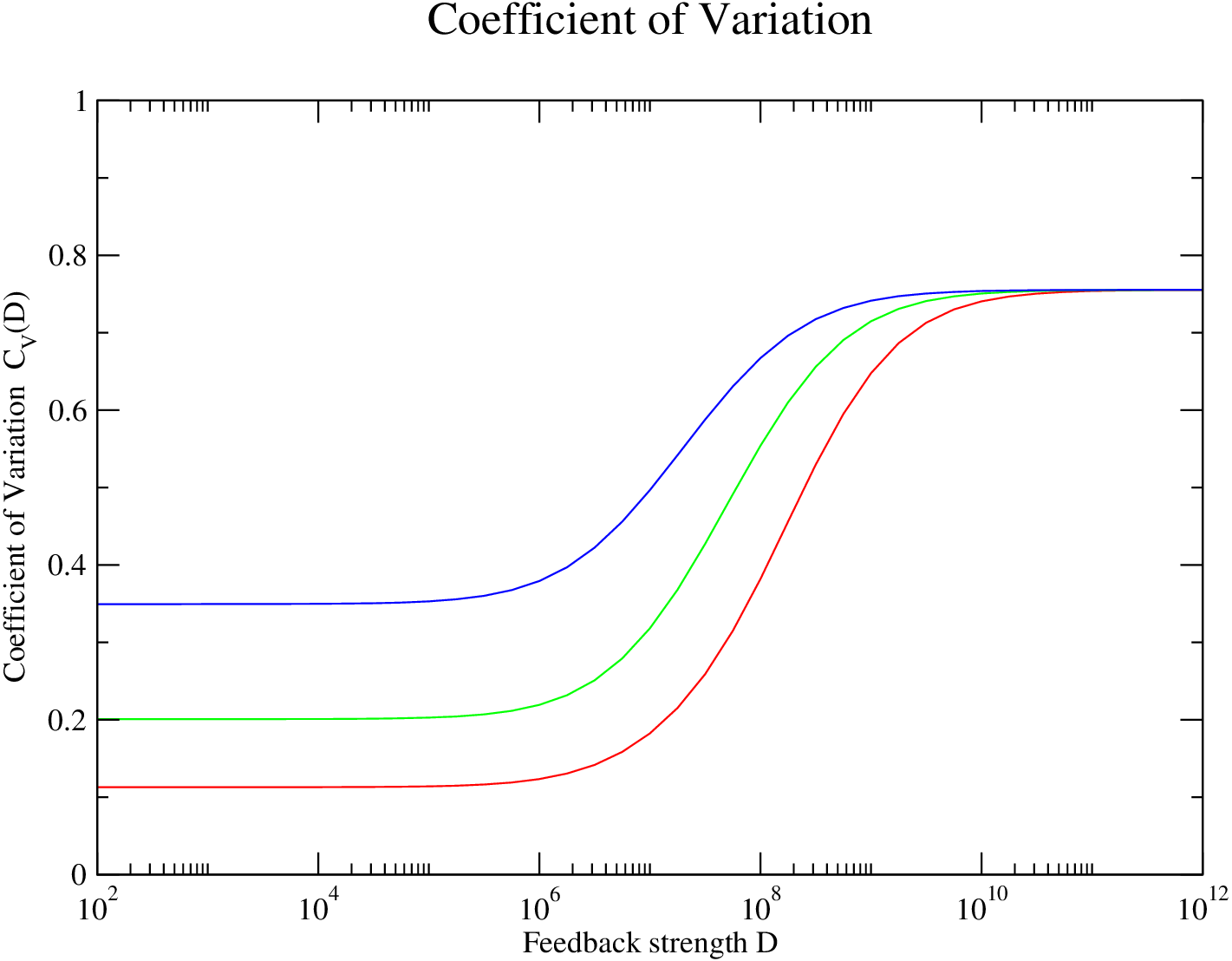}}
\end{picture} 
\caption{\small{
We show the coefficient of variation (standard deviation/mean). Horizontal 
axis denotes feedback strength $D$ and vertical axis denotes the coefficient of variation $C_V$ (standard deviation/mean). The numerical values of the system parameters are the same as Fig. \ref{fig: expectation}. but noise size $\sigma$ takes the following values. 
red curve $\sigma=1.5\times 10^{-7} [\mbox{Ms}^{-1/2}]$, green curve $\sigma=1.5\times 10^{-6.75}[\mbox{Ms}^{-1/2}]$, and blue curve $\sigma=1.5 \times 10^{-6.5} [\mbox{Ms}^{-1/2}]$. In the strong feedback region, all three curves go to the same value 0.755}}
\label{fig: adjusting time}
\end{figure}

\begin{figure}[htb]
\setlength{\unitlength}{1cm}
\begin{picture}(15,12)(-1,-1)
\put(-1,0){\includegraphics[scale=0.5]{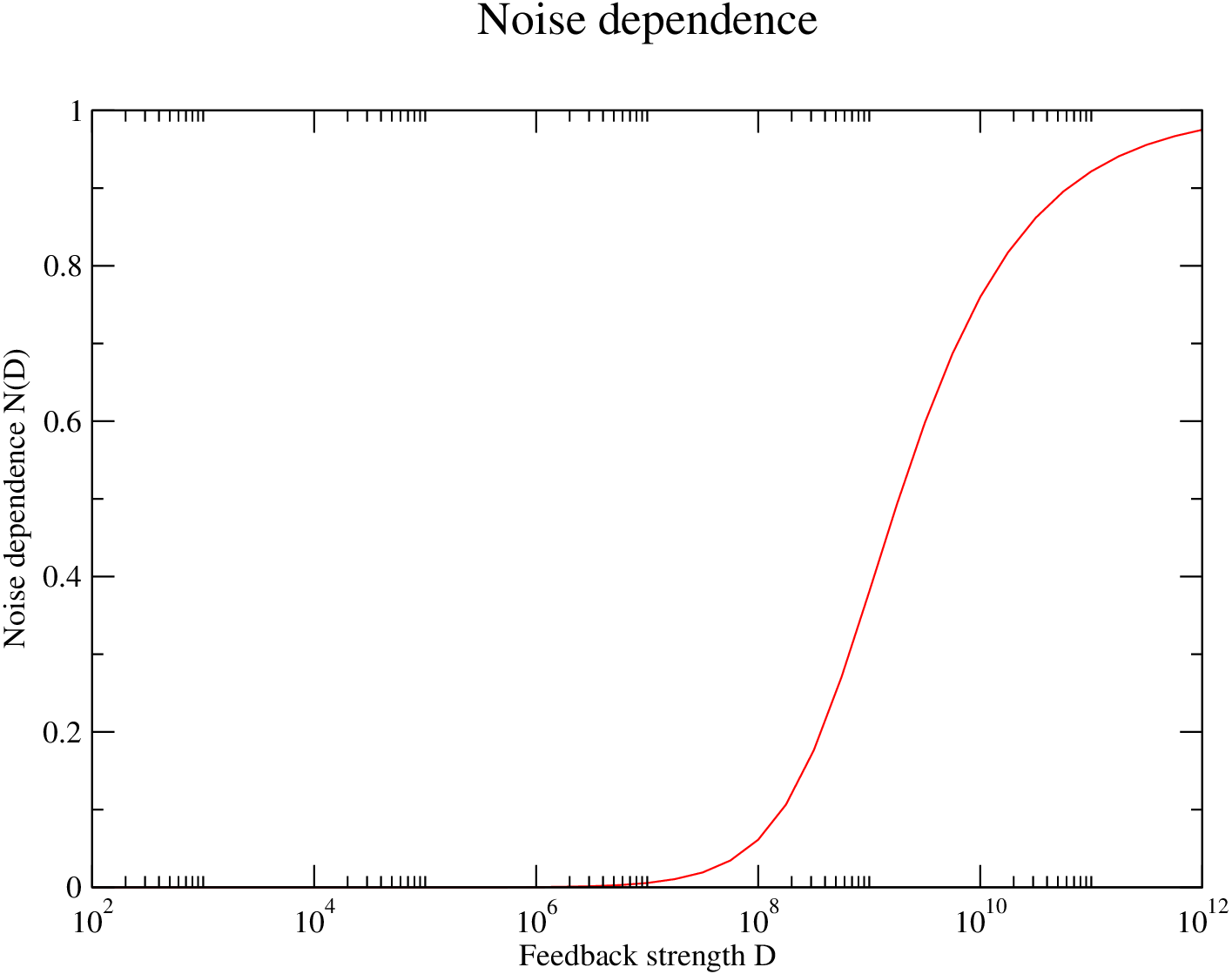}}
\end{picture} 
\caption{\small{
We show the noise dependence $N(D)$. Horizontal axis denotes feedback strength $D$ and vertical axis denotes the noise dependence $N(D)$. The numerical values of the system parameters are the same as Fig. \ref{fig: expectation}.}}
\label{fig: adjusting time}
\end{figure}


\begin{thebibliography}{99}

\bibitem{kaern} Kaern, M. T.C. Elston, W.J. Blake, and J.J. Collins, 2005. Stochasticity in gene expression; from theories to phenotypes.
Reviews Genetics 6, 451-464.

\bibitem{swain2002} Swain, P.S., Elowitz M.B. and Siggia E.D. (2002). Intrinsic and extrinsic contributions to stochasticity in gene
expression. Proc. Natl. Acad. Sci. USA 99, 12795-12800.

\bibitem{kepler2001} Kepler T.B. and Elston TC (2001) Stochasticity in transcriptional regulation: origins, consequences and mathematical representations. Biophys. J. 81, 3116-3136.

\bibitem{chen2005} Chen, L., Wang, R., Zhou T. and Aihara K. (2005). Noise-induced cooperative behavior in multicell system. Bioinformatics
21, 2722-2535.

\bibitem{gardner} Gardner, T.S., C.R. Cantor and J.J. Collins, 2000. Construction of a genetic toggle switch in E. Coli. 
Nature 403, 339-342.

\bibitem{paulsson2004} Paulsson J (2004). Summing up the noise in gene networks. Nature 427, 415-418.

\bibitem{elowitz2002} Elowitz M.B., Levine A.J., Siggia E.D., Swain P.S. (2002). Stochastic gene expression in a single cell. Science 297,
1183-1186.

\bibitem{austin2006} Austin D.W., Allen M.S., McCollum J.M., Dar RD, Wilgus J.R. and Simpson M.L. (2006) Gene network shaping of inherent
noise spectra. Nature 439, 608-611.


\bibitem{blake2003} Blake W.J., Kaern M, Cantor C.E., Collins J.J. (2003) Noise in eukaryotic gene expression. Nature 422, 622-637.

\bibitem{raser2004} Raser J.M., OfShea E.K. (2004) Control of stochasticity in eukaryotic gene expression. Science 304:1811-1814.

\bibitem{savageau1974} Savageau MA (1974) Comparison of classical and autogenous systems of regulation in inducible operons. Nature 252: 546-549

 
\bibitem{becskei} Becskei, A. and L. Serrano, 2000. Engineering stability in gene networks by autoregulation. Nature 405, 590-593.


\bibitem{rosenfeld2002} Rosenfeld, N., M. B. Elowitz, and U. Alon. 2002. Negative autoregulation speeds the response times of transcription networks. J. Mol. Biol. 323:785?793.


\bibitem{pedraza2005} Pedraza J.M. and van Oudenaarden A. (2005). Noise propagation in gene networks. Science 307:1965-1969.

\bibitem{hooshangi2005} Hooshangi S., Thiberge S., Weiss R. (2005) Ultrasensitivity and noise propagation in a synthetic transcriptional cascade. Proc. Natl. Acad. Sci. USA 102:3581-3586.

\bibitem{serrano2} Dublanche Y., Michalodimitrakis K., Kummerer N., Foglierini M. And Serrano L. (2006). Mol. Systems Biology 41, 1-12.




\bibitem{Denise} Wolf D.M. and F.H. Eeckman, 1998. On the relationship between genomic regulatory element orgnization and gene regulatory
dynamics. J. Theor. Biol. 195, 167-186.
 
\bibitem{Wong} Wong, E., 1971. {\it Stochastic Processes in Information and Dynamical 
Systems}, Ed. New York, McGraw-Hill.
\bibitem{Kampen} van Kampen, N.G., 1992. {\it Stochastic processes in physics and chemistry,} Elsevier Science B.V.
\bibitem{black} Mikosch, T., 1998. {\it Elementary Stochastic Calculus with Finance in View}, World Scientific Publishing Co. Pte. Ltd.




\bibitem{ochi} T. Ochiai, J.C. Nacher, T. Akutsu, Physics Letters A 330 
(2004) pp.313.

\bibitem{ochi2} T. Ochiai, J.C. Nacher, T. Akutsu, Physics Letters A, 339 
(2005) pp.1.
 












 



\end{thebibliography}
\end{document}